\documentclass[intlimits,twoside,a4paper]{article}

\usepackage[cp1251]{inputenc}
\usepackage{xcolor}
\usepackage[eqsecnum]{cmpj3}

\usepackage{bm}

\issue{2020}{23}{4}{43708}
\doinumber{10.5488/CMP.23.43708}
\title[Nature of a magnetization plateau of~3$d$-4$f$ coordination polymer (Dy$_2$Cu$_2$)$_n$]
{Insights into nature of a magnetization plateau of~3$d$-4$f$ coordination polymer [Dy$_2$Cu$_2$]$_n$ from a~spin-1/2 Ising-Heisenberg orthogonal-dimer chain}
\author[J. Stre\v{c}ka, L. G\'alisov\'a, T. Verkholyak]{J. Stre\v{c}ka\refaddr{upjs}, L. G\'alisov\'a\refaddr{tuke}, T. Verkholyak\refaddr{icmp}}
\addresses{
\addr{upjs} Department of Theoretical Physics and Astrophysics, Faculty of Science, P.~J.~\v{S}af\'arik University, \\Park Angelinum 9, 040\,01 Ko\v{s}ice, Slovakia
\addr{tuke} Institute of Manufacturing Management, Faculty of Manufacturing Technologies with the seat in Pre\v{s}ov, \\
            Technical University of~Ko\v{s}ice, Bayerova 1, 080\,01 Pre\v{s}ov, Slovakia \\
\addr{icmp} Institute for Condensed Matter Physics of the National
Academy of Sciences of Ukraine,\\ 1 Svientsitskii St., 79011 Lviv,
Ukraine}
\date{Received June 22, 2020, in final form August 13, 2020}
\begin{document}

\maketitle

\begin{abstract}
The ground state and magnetization process of an exactly solved spin-$1/2$ Ising-Heisenberg orthogonal-dimer chain with two different gyromagnetic factors of the Ising and Heisenberg spins are investigated in detail. It is shown that the investigated quantum spin chain exhibits up to  seven possible ground states depending on a mutual interplay of the magnetic field, intra- and inter-dimer coupling constants. More specifically, the frustrated and modulated quantum antiferromagnetic phases are responsible in zero-temperature magnetization curves for a zero magnetization plateau. The intermediate 1/11- and 5/11-plateaus emerge due to the frustrated and modulated quantum ferrimagnetic phases, while the intermediate 9/11- and 10/11-plateaus can be attributed to the quantum and classical ferrimagnetic phases. It is conjectured that the magnetization plateau experimentally observed in a high-field magnetization curve of 3$d$-4$f$ heterobimetallic coordination polymer [\{Dy(hfac)$_2$(CH$_3$OH)\}$_2$\{Cu(dmg)(Hdmg)\}$_2$]$_n$ (H$_2$dmg $=$ dimethylglyoxime; Hhfac $=$ 1,1,1,5,5,5-hexafluoropentane-2,4-dione) could be attributed to the classical and quantum ferrimagnetic phases. 

\keywords Ising-Heisenberg orthogonal-dimer chain, magnetization plateau, 3$d$-4$f$ coordination polymer
\end{abstract}

\section{Introduction}
\label{sec:1}

The frustrated spin-$\frac{1}{2}$ Heisenberg orthogonal-dimer (or dimer-plaquette) chain~\cite{Iva97,Rich98,Kog00,Miy11} has attracted considerable attention as it represents one-dimensional counterpart of the famous Shastry-Sutherland model \cite{Sha81}, which is widely studied by virtue of elucidation a peculiar sequence of fractional plateaus experimentally observed in low-temperature magnetization curves of SrCu$_2$(BO$_3$)$_2$ \cite{Mas13} and rare-earth tetraborides RB$_4$ \cite{Gab20}. It has been argued by Schulenburg and Richter \cite{Sch02a,Sch02b} that a zero-temperature magnetization curve of the spin-$\frac{1}{2}$ Heisenberg orthogonal-dimer chain displays an infinite series of fractional magnetization plateaus at rational numbers $\frac{n}{2n+2}=\frac{1}{4}, \frac{1}{3}, \dots, \frac{1}{2}$, whereas the lowermost and uppermost plateaus from this series are the widest ones. Another interesting feature of the spin-$\frac{1}{2}$ Heisenberg orthogonal-dimer chain lies in its belonging to a prominent class of flat-band models, for which low-temperature thermodynamics can be elaborated by making use of an effective lattice-gas description as extensively discussed by Derzhko and coworkers \cite{Der06,Der15}.

Regrettably, thermodynamic properties of quantum Heisenberg spin models are in general inaccessible by exact calculations at nonzero temperature. Replacement of some of the quantum Heisenberg spins by the classical Ising ones paves the way to an exact solution of the analogous Ising-Heisenberg models by employing exact mapping transformations and the transfer-matrix method \cite{Fis59,Roj09,Str10}. For instance a few exactly solved versions of the spin-$\frac{1}{2}$ Ising-Heisenberg orthogonal-dimer chain \cite{Oha12,Pau13,Ver13,Ver14,Gal20}, to be further abbreviated as IH-ODC,  brought a deeper insight into the magnetization process \cite{Oha12,Ver13,Gal20}, magnetocaloric effect \cite{Oha12,Ver13}, low-temperature thermodynamics \cite{Pau13,Ver13,Ver14} and bipartite thermal entanglement \cite{Pau13,Gal20} of this frustrated quantum spin chain. Besides, it was demonstrated that the exact solution for the IH-ODC may serve as a useful starting point for the many-body perturbation treatment of the fully Heisenberg counterpart model within a more advanced strong-coupling approach \cite{Ver16}.

\begin{figure}[!t]
\centerline{\includegraphics[width=0.99\textwidth]{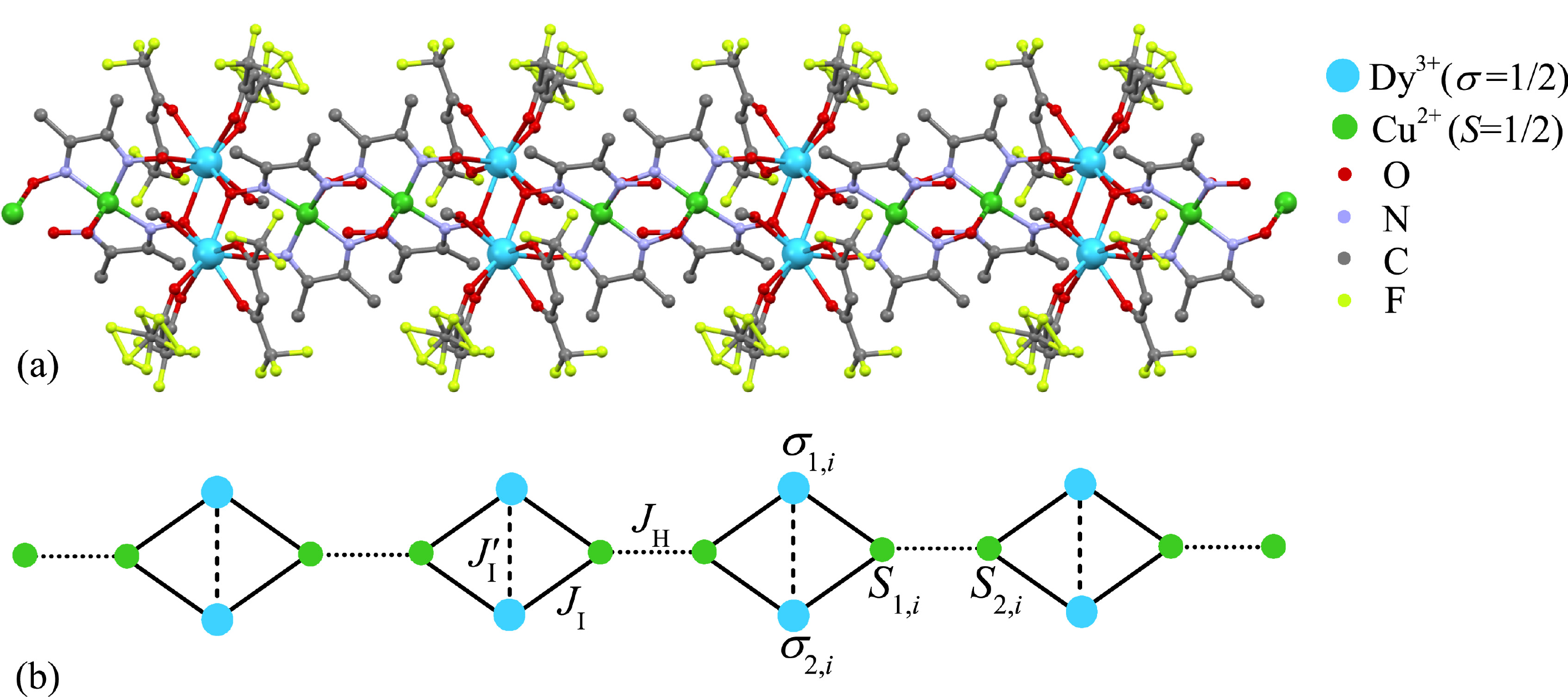}}
\caption{(Colour online) (a) A crystal structure of 3$d$-4$f$ coordination polymer [Dy$_2$Cu$_2$]$_n$ (see the text for a full chemical formula) adapted according to crystallographic data reported in reference \cite{Oka08}. Large cyan balls determine crystallographic positions of Dy$^{3+}$ magnetic ions, while small green balls stand for crystallographic positions of Cu$^{2+}$ magnetic ions (a coloured scheme for atom labeling is presented in the legend); (b) The magnetic structure of the corresponding IH-ODC, in which Dy$^{3+}$ magnetic ions are treated as the Ising spins while Cu$^{2+}$ magnetic ions are treated as the Heisenberg spins. The coupling constants $J_\text{I}$, $J_\text{I}^{\prime}$ and $J_\text{H}$ are assigned to the Ising inter-dimer interaction between  Dy$^{3+}$ and Cu$^{2+}$ magnetic ions (solid lines), the Ising intra-dimer interaction between Dy$^{3+}$ magnetic ions (dashed lines) and the Heisenberg intra-dimer interaction between Cu$^{2+}$ magnetic ions (dotted lines), respectively.} 
\label{fig1}
\end{figure}

Although we are currently not aware of any experimental realization of the spin-$\frac{1}{2}$ Heisenberg orthogonal-dimer chain, it surprisingly turns out that 3$d$-4$f$ heterobimetallic coordination polymer [\{Dy(hfac)$_2$(CH$_3$OH)\}$_2$\{Cu(dmg)(Hdmg)\}$_2$]$_n$ (H$_2$dmg $=$ dimethylglyoxime; Hhfac $=$ 1,1,1,5,5,5-hexafluoropentane-2,4-dione) \cite{Oka08}, hereafter abbreviated as [Dy$_2$Cu$_2$]$_n$, provides an experimental realization of the IH-ODC. In fact, the polymeric compound  [Dy$_2$Cu$_2$]$_n$ displays a peculiar one-dimensional architecture with regularly alternating dimeric units of Dy$^{3+}$-Dy$^{3+}$ and Cu$^{2+}$-Cu$^{2+}$ magnetic ions stacked in an orthogonal fashion with respect to each other as illustrated in figure~\ref{fig1}~(a) \cite{Oka08}. It should be pointed out, moreover, that Dy$^{3+}$ magnetic ion represents Kramers ion with the ground-state multiplet $^{6}$H$_{15/2}$, which is subjected to a relatively strong crystal-field splitting into eight well-separated Kramers doublets~\cite{Jon74,Jen91}. In this regard, the magnetic behaviour of Dy$^{3+}$ magnetic ion with the total angular momentum $J=15/2$ and the associated $g$-factor $g_J = 4/3$ can be approximated at low enough temperatures by the classical Ising spin with the effective gyromagnetic factor $g_\text{Dy} = 20$ when neglecting the admixture of all excited Kramers dublets \cite{Jon74,Jen91}. Getting back to a magnetic structure of the polymeric complex [Dy$_2$Cu$_2$]$_n$, which is schematically drawn in figure~\ref{fig1}~(b), the vertical dimer of Dy$^{3+}$-Dy$^{3+}$ magnetic ions may be approximated by a couple of the Ising spins, while the horizontal dimer of Cu$^{2+}$-Cu$^{2+}$ magnetic ions may be approximated by a couple of the Heisenberg spins.  

The organization of this article is as follows. In section \ref{sec:2} we describe the studied IH-ODC and recall basic steps of its exact analytical solution. The ground state and magnetization process of the investigated quantum spin chain are theoretically studied in section \ref{sec:3}. The available experimental data for the high-field magnetization curve of the polymeric compound [Dy$_2$Cu$_2$]$_n$ are interpreted by virtue of the IH-ODC in section \ref{sec:4}. Finally, several conclusions and future outlooks are mentioned in section \ref{sec:5}.
\newpage

\section{Spin-$\frac{1}{2}$ Ising-Heisenberg orthogonal-dimer chain}
\label{sec:2}

Let us start by introducing the IH-ODC in a magnetic field through the following Hamiltonian [see figure~\ref{fig1}~(b) for a schematic illustration]:
\begin{eqnarray}
\label{eq:Htot}
\hat{\cal H} \!\!\!&=&\!\!\!
J_\text{H} \sum_{i = 1}^{N} \hat{\mathbf S}_{1,i}\cdot\hat{\mathbf S}_{2,i} + 
J_\text{I}^{\prime} \sum_{i = 1}^{N} \hat{\sigma}_{1,i}^{z}\hat{\sigma}_{2,i}^{z} 
+ J_\text{I} \sum_{i = 1}^{N} \big[\hat{S}_{1,i}^{z}\big(\hat{\sigma}_{1,i}^{z}\!+\hat{\sigma}_{2,i}^{z}\big) +
\hat{S}_{2,i}^{z}\big(\hat{\sigma}_{1,i+1}^{z}\!+\hat{\sigma}_{2,i+1}^{z}\big)\big] \nonumber \\ 
\!\!\!&-&\!\!\!\! 
g_\text{H}\mu_\text{B}B\sum_{i = 1}^{N} \big(\hat{S}_{1,i}^{z} \!+ \hat{S}_{2,i}^{z}\big)
- g_\text{I}\mu_\text{B}B\sum_{i = 1}^{N} \big(\hat{\sigma}_{1,i}^{z}\!+\hat{\sigma}_{2,i}^{z}\big),
\end{eqnarray}
where $\hat{\mathbf S}_{1(2),i}\equiv (\hat{S}_{1(2),i}^{x}, \hat{S}_{1(2),i}^{y}, \hat{S}_{1(2),i}^{z})$ denote standard spin-$\frac{1}{2}$ operators ascribed to Cu$^{2+}$ magnetic ions approximated by the notion of quantum Heisenberg spins and $\hat{\sigma}_{1(2),i}^z$ refer to the $z$-component of the standard spin-$\frac{1}{2}$ operators ascribed to Dy$^{3+}$ magnetic ions approximated by the notion of classical Ising spins. The coupling constants $J_\text{H}$ and $J_\text{I}^{\prime}$ determine a strength of the Heisenberg and Ising intra-dimer interactions within the horizontal Cu$^{2+}$-Cu$^{2+}$ and vertical Dy$^{3+}$-Dy$^{3+}$ dimers, respectively, while the coupling constant $J_\text{I}$ determines a strength of the Ising inter-dimer interaction between the nearest-neighbour Cu$^{2+}$ and  Dy$^{3+}$ magnetic ions. The Zeeman's terms $h_\text{H} = g_\text{H}\mu_\text{B}B$ and $h_\text{I} = g_\text{I}\mu_\text{B}B$ take into account a magnetostatic energy of magnetic moments relating to the Heisenberg and Ising spins in presence of the external magnetic field $B$, which differ due to different gyromagnetic factors $g_\text{H}$ and $g_\text{I}$ of Cu$^{2+}$ and  Dy$^{3+}$ magnetic ions, respectively. 

It is worthwhile to remark that the partition function, Gibbs free energy and magnetization of the IH-ODC defined through the Hamiltonian \eqref{eq:Htot} was exactly calculated under the periodic boundary conditions $\hat{\sigma}_{1(2),N+1}^{z} \equiv \hat{\sigma}_{1(2),1}^{z}$ in our preceding paper \cite{Gal20}. The  calculation procedure used takes advantage of splitting the total Hamiltonian \eqref{eq:Htot} into commuting six-spin cluster Hamiltonians involving one horizontal Cu$^{2+}$-Cu$^{2+}$ dimer and two enclosing vertical Dy$^{3+}$-Dy$^{3+}$ dimers, which allow a straightforward factorization of the partition function into a product of the respective Boltzmann factors. After tracing out spin degrees of freedom of the horizontal Cu$^{2+}$-Cu$^{2+}$ dimer (Heisenberg dimer), the partition function is in fact expressed in terms of four-by-four transfer matrix depending on spin states of two adjacent vertical Dy$^{3+}$-Dy$^{3+}$ dimers (Ising dimers) and the whole magnetothermodynamics can be elaborated by making use of the transfer-matrix method (the readers interested in further calculation details are referred to reference \cite{Gal20}). However, all numerical results presented in reference \cite{Gal20} were restricted just to the particular case $h_\text{H} = h_\text{I}$, which corresponds to setting the same Land\'e $g$-factors for Cu$^{2+}$ and  Dy$^{3+}$ magnetic ions which is contrary to the expected (typical) values of the gyromagnetic factors $g_\text{H} \approx 2$ for Cu$^{2+}$ magnetic ions and $g_\text{I} \approx 20$ for Dy$^{3+}$ magnetic ions. Therefore, in the present article we  adapt the exact solution for the IH-ODC reported in reference \cite{Gal20} in order to investigate the effect of different `local magnetic fields' $h_\text{H} \neq h_\text{I}$ arising from the difference of the gyromagnetic factors of Cu$^{2+}$ and Dy$^{3+}$ magnetic ions 
$g_\text{H} \neq g_\text{I}$.

\section{Theoretical results}
\label{sec:3}

In this section we  examine in detail the ground state and magnetization process of the IH-ODC by assuming the gyromagnetic factors $g_\text{H} = 2$ and $g_\text{I} = 20$, which are close to typical values of the Land\'e $g$-factors for Cu$^{2+}$ and  Dy$^{3+}$ magnetic ions, respectively. To reduce the number of free parameters, a size of the coupling constant $J_\text{I} >0$ corresponding to the antiferromagnetic Ising inter-dimer interaction between Dy$^{3+}$-Cu$^{2+}$ magnetic ions  serves as an energy unit when defining a relative strength of the Heisenberg intra-dimer interaction $J_\text{H}/J_\text{I}$ within the horizontal dimers, the Ising intra-dimer interaction $J_\text{I}^{\prime}/J_\text{I}$ within the vertical dimers and the magnetic field $\mu_\text{B}B/J_\text{I}$.  

\subsection{Ground state}
\label{subsec:3.1}

\begin{figure}[!b]
\vspace{-0.2cm}
\centerline{\includegraphics[width=0.92\textwidth]{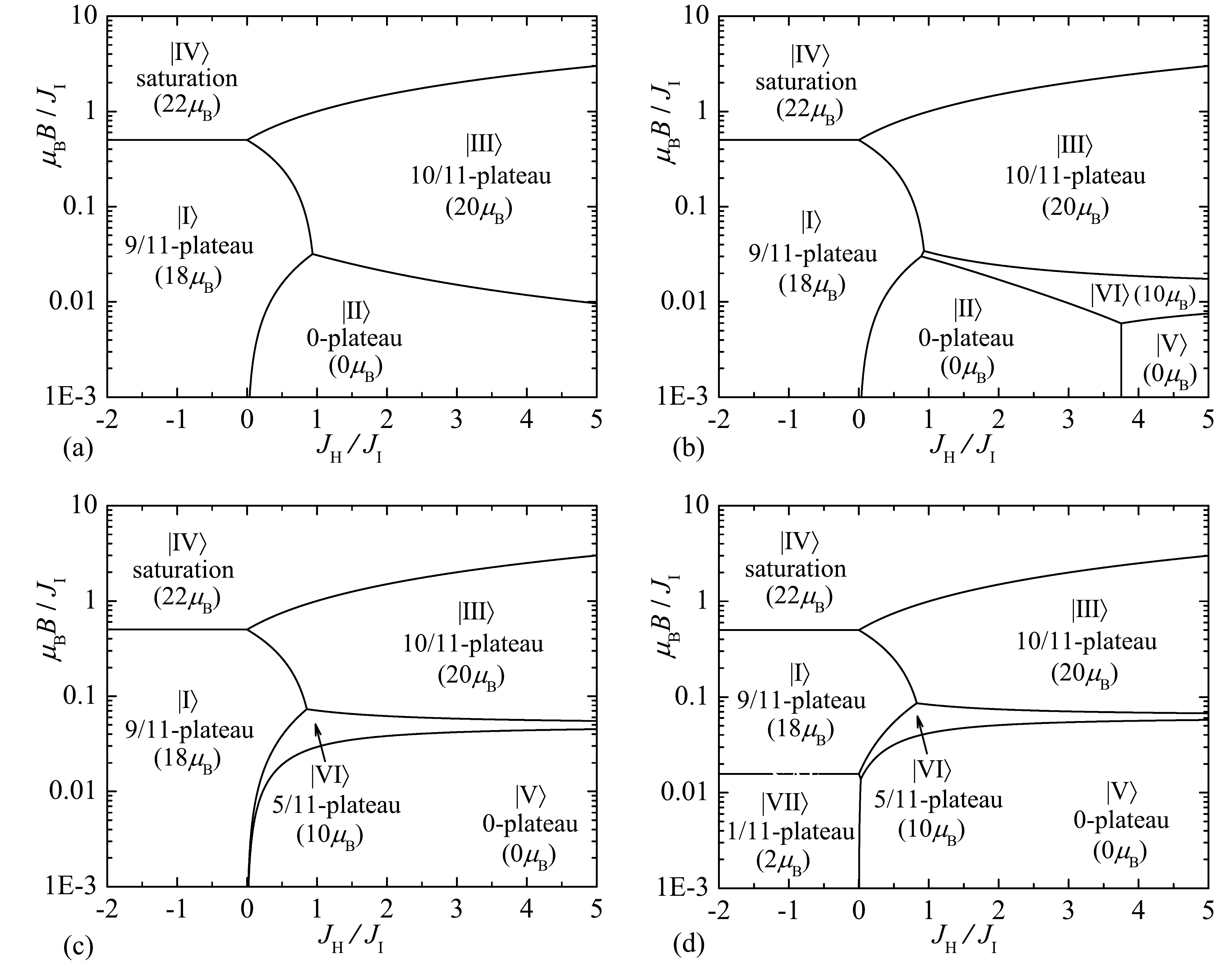}}
\vspace{-0.25cm}
\caption{A semi-logarithmic plot of the ground-state phase diagram of the IH-ODC with the gyromagnetic factors $g_\text{H} = 2$ and $g_\text{I} = 20$ in the $J_\text{H}/J_\text{I}{-}\mu_\text{B}B/J_\text{I}$ plane for four representative values of the interaction ratio: (a) $J_\text{I}^{\prime}/J_\text{I} = -0.5$; (b) $J_\text{I}^{\prime}/J_\text{I} = 0.5$; (c) $J_\text{I}^{\prime}/J_\text{I} = 2.0$; (d) $J_\text{I}^{\prime}/J_\text{I} = 2.5$. The numbers in round brackets determine the total magnetization in units of Bohr magneton $\mu_\text{B}$ and the fractions represent its relative size with respect to the saturation magnetization.} 
\label{fig2}
\end{figure}

The IH-ODC with the gyromagnetic factors $g_\text{H} = 2$ and $g_\text{I} = 20$ may display, in presence of the magnetic field, up to seven different ground states depending on a mutual interplay of the coupling constants $J_\text{H}/J_\text{I}$, $J_\text{I}^{\prime}/J_\text{I}$ and the magnetic field $\mu_\text{B}B/J_\text{I}$. The typical ground-state phase diagrams are reported in figure~\ref{fig2} in the $J_\text{H}/J_\text{I}{-}\mu_\text{B}B/J_\text{I}$ parameter plane for four representative values of the interaction ratio  $J_\text{I}^{\prime}/J_\text{I} = -0.5$, $0.5$, $2.0$ and $2.5$. It is quite evident from figure~\ref{fig2}~(a) that the ground-state phase diagram for the particular case with the ferromagnetic Ising intra-dimer coupling $J_\text{I}^{\prime}/J_\text{I} = -0.5$ involves just four different ground states, more specifically, the classical ferrimagnetic phase $|\text{I}\rangle$ with an antiparallel spin arrangement of the Ising and Heisenberg dimers characterized through the following eigenvector and corresponding eigenenergy
\begin{eqnarray}
\label{eq:I}
|\text{I}\rangle = \prod_{i=1}^{N}
{\textstyle\left|{\uparrow\atop\uparrow}\right\rangle_{\!i}}
\otimes
|\!\downarrow\downarrow\rangle_{i}\,, \qquad
E_{\text{I}} = \frac{N}{4}\left(J_\text{H} + J_\text{I}^{\prime} - 4J_\text{I} - 4h_\text{I} + 4h_\text{H}\right), 
\end{eqnarray}	
the modulated quantum antiferromagnetic phase $|\text{II}\rangle$ with alternating character of the Ising dimers and a singlet-like state of the Heisenberg dimers  
\begin{eqnarray}
\label{eq:II}
|\text{II}\rangle \!\!\!&=&\!\!\! \prod_{i=1}^{N/2}
{\textstyle\left|{\uparrow\atop\uparrow}\right\rangle_{\!2i-1}}
\otimes\Big(
\sin\varphi_{2}|\!\uparrow\downarrow\rangle_{2i-1} - \cos\varphi_{2}|\!\downarrow\uparrow\rangle_{2i-1}
\Big)
\otimes\!
\left.\big|\mbox{\normalsize${\downarrow\atop\downarrow}$}\right\rangle_{\!2i} \otimes\Big(
\cos\varphi_{2}|\!\uparrow\downarrow\rangle_{2i} - \sin\varphi_{2}|\!\downarrow\uparrow\rangle_{2i}
\Big),
\nonumber\\
E_{\text{II}} \!\!\!&=&\!\!\! -\frac{N}{4}\left(J_\text{H} + 2\!\sqrt{4J_\text{I}^{2} +J_\text{H}^{2}} - J_\text{I}^{\prime}\right), \qquad \qquad \varphi_2 = \frac{1}{2} \arctan \left(\frac{J_\text{H}}{2 J_\text{I}} \right),
\end{eqnarray}	
the quantum ferrimagnetic phase $|\text{III}\rangle$ with fully polarized Ising dimers and a perfect singlet-dimer state of the Heisenberg dimers
\begin{eqnarray}
\label{eq:III}
|\text{III}\rangle = \prod_{i=1}^{N}
{\textstyle\left|{\uparrow\atop\uparrow}\right\rangle_{\!i}}
\otimes
\dfrac{1}{\sqrt{2}}
\left(
|\!\uparrow\downarrow\rangle_{i} - |\!\downarrow\uparrow\rangle_{i}
\right), \qquad
E_{\text{III}} = -\frac{N}{4}\left(3J_\text{H} - J_\text{I}^{\prime} + 4h_\text{I}\right),
\end{eqnarray}	
and finally, the saturated paramagnetic phase $|\text{IV}\rangle$ with a fully polarized nature of the Ising as well as Heisenberg dimers
\begin{eqnarray}
\label{eq:IV}
|\text{IV}\rangle = \prod_{i=1}^{N}
{\textstyle\left|{\uparrow\atop\uparrow}\right\rangle_{\!i}}
\otimes
|\!\uparrow\uparrow\rangle_{i}\,, \qquad
E_{\text{IV}} = \frac{N}{4}\left(J_\text{H} + J_\text{I}^{\prime} + 4J_\text{I} - 4h_\text{I} - 4h_\text{H}\right).
\end{eqnarray}
Note that all eigenvectors are written as a tensor product over states of the vertical Ising dimers (former state vector) and the horizontal Heisenberg dimers (the latter state vector), respectively. At low enough magnetic fields, the classical ferrimagnetic phase $|\text{I}\rangle$ dominates in the parameter region with the ferromagnetic Heisenberg intra-dimer coupling $J_\text{H}<0$, while the modulated quantum antiferromagnetic phase $|\text{II}\rangle$ and the quantum ferrimagnetic phase $|\text{III}\rangle$ dominate in the parameter region with the antiferromagnetic Heisenberg intra-dimer coupling $J_\text{H}>0$. Of course, the saturated paramagnetic phase $|\text{IV}\rangle$ represents the actual ground state at high enough magnetic fields regardless of whether the ferromagnetic or antiferromagnetic Heisenberg intra-dimer coupling is considered.

On the other hand, the ground-state phase diagram of the IH-ODC with a relatively weak antiferromagnetic Ising intra-dimer coupling  additionally involves the other two ground states [see figure~\ref{fig2}~(b) for $J_\text{I}^{\prime}/J_\text{I} = 0.5$], which could be identified as the frustrated quantum antiferromagnetic phase $|\text{V}\rangle$ with a two-fold degenerate antiferromagnetic state of the Ising dimers and a perfect singlet-dimer state of the Heisenberg dimers 
\begin{eqnarray}
\label{eq:V}
|\text{V}\rangle = \prod_{i=1}^{N}
{\textstyle\left|{\uparrow\atop\downarrow}\right\rangle_{\!i}}\,
\left(\text{or}\left.\big|\mbox{\normalsize${\downarrow\atop\uparrow}$}\right\rangle_{\!i}\right)\otimes
\dfrac{1}{\sqrt{2}}
\left(
|\!\uparrow\downarrow\rangle_{i} - |\!\downarrow\uparrow\rangle_{i}
\right),
\quad
E_\text{V} = -\frac{N}{4}\left(3J_\text{H} + J_\text{I}^{\prime}\right),
\end{eqnarray}	
and the highly degenerate modulated quantum ferrimagnetic phase $|\text{VI}\rangle$ with alternating ferro-antiferromagnetic character of the Ising dimers and a singlet-like state of the Heisenberg dimers  
\begin{eqnarray}
\label{eq:VI}
|\text{VI}\rangle \!\!\!&=&\!\!\! \prod_{i=1}^{N/2}
{\textstyle\left|{\uparrow\atop\uparrow}\right\rangle_{\!2i-1}}\!\!
\otimes\!\Big(
\sin\varphi_{6}|\!\uparrow\downarrow\rangle_{2i-1} - \cos\varphi_{6}|\!\downarrow\uparrow\rangle_{2i-1}
\Big)
\otimes\!
\left.\big|\mbox{\normalsize${\uparrow\atop\downarrow}$}\right\rangle_{\!2i} \!\!
\,\left(\text{or}\left.\big|\mbox{\normalsize${\downarrow\atop\uparrow}$}\right\rangle_{\!2i}\right)\!\otimes\!\Big(
\cos\varphi_{6}|\!\uparrow\downarrow\rangle_{2i} - \sin\varphi_{6}|\!\downarrow\uparrow\rangle_{2i}
\Big),
\nonumber\\
E_{\text{VI}} \!\!\!&=&\!\!\! -\frac{N}{4}\left(J_\text{H} + 2\!\sqrt{J_\text{I}^{2} +J_\text{H}^{2}} + 2h_\text{I}\right), \qquad  \qquad \varphi_6 = \frac{1}{2} \arctan \left(\frac{J_\text{H}}{J_\text{I}} \right). 
\end{eqnarray}
It is quite clear from figure~\ref{fig2}~(b) that the ground-state phase diagram has not  changed in the parameter space with the ferromagnetic Heisenberg intra-dimer coupling $J_\text{H}<0$, while two novel ground states emerge at low (up to moderate) magnetic fields in the parameter space with sufficiently strong antiferromagnetic Heisenberg intra-dimer coupling $J_\text{H}>0$. It is noteworthy that the frustrated quantum antiferromagnetic phase $|\text{V}\rangle$ suppresses the modulated quantum antiferromagnetic phase $|\text{II}\rangle$ upon increasing the relative strength of the antiferromagnetic Ising intra-dimer coupling $J_\text{I}^{\prime}/J_\text{I}$ until this latter ground state completely vanishes from the ground-state phase diagram as evidenced by figure~\ref{fig2}~(c) for $J_\text{I}^{\prime}/J_\text{I} = 2.0$. Last but not least, the sufficiently strong antiferromagnetic Ising intra-dimer coupling $J_\text{I}^{\prime}/J_\text{I}>2$ may additionally cause an uprise of the novel ground state also in the parameter space with the ferromagnetic Heisenberg intra-dimer coupling $J_\text{H}<0$ and small enough magnetic fields, which could be identified as the frustrated ferrimagnetic phase $|\text{VII}\rangle$ with a two-fold degenerate antiferromagnetic state of the Ising dimers and fully polarized Heisenberg dimers given by
\begin{eqnarray}
\label{eq:VII}
|\text{VII}\rangle = \prod_{i=1}^{N}
{\textstyle\left|{\uparrow\atop\downarrow}\right\rangle_{\!i}}\,
\left(\text{or}\left.\big|\mbox{\normalsize${\downarrow\atop\uparrow}$}\right\rangle_{\!i}\right)\otimes
|\!\uparrow\uparrow\rangle_{i},
\qquad
E_{\text{VII}} = \frac{N}{4}\left(J_\text{H} - J_\text{I}^{\prime} - 4h_\text{H}\right).
\end{eqnarray}

\subsection{Zero- and low-temperature magnetization curves}
\label{subsec:3.2}
\begin{figure}[!b]
\vspace{-0.3cm}
\centerline{\includegraphics[width=0.92\textwidth]{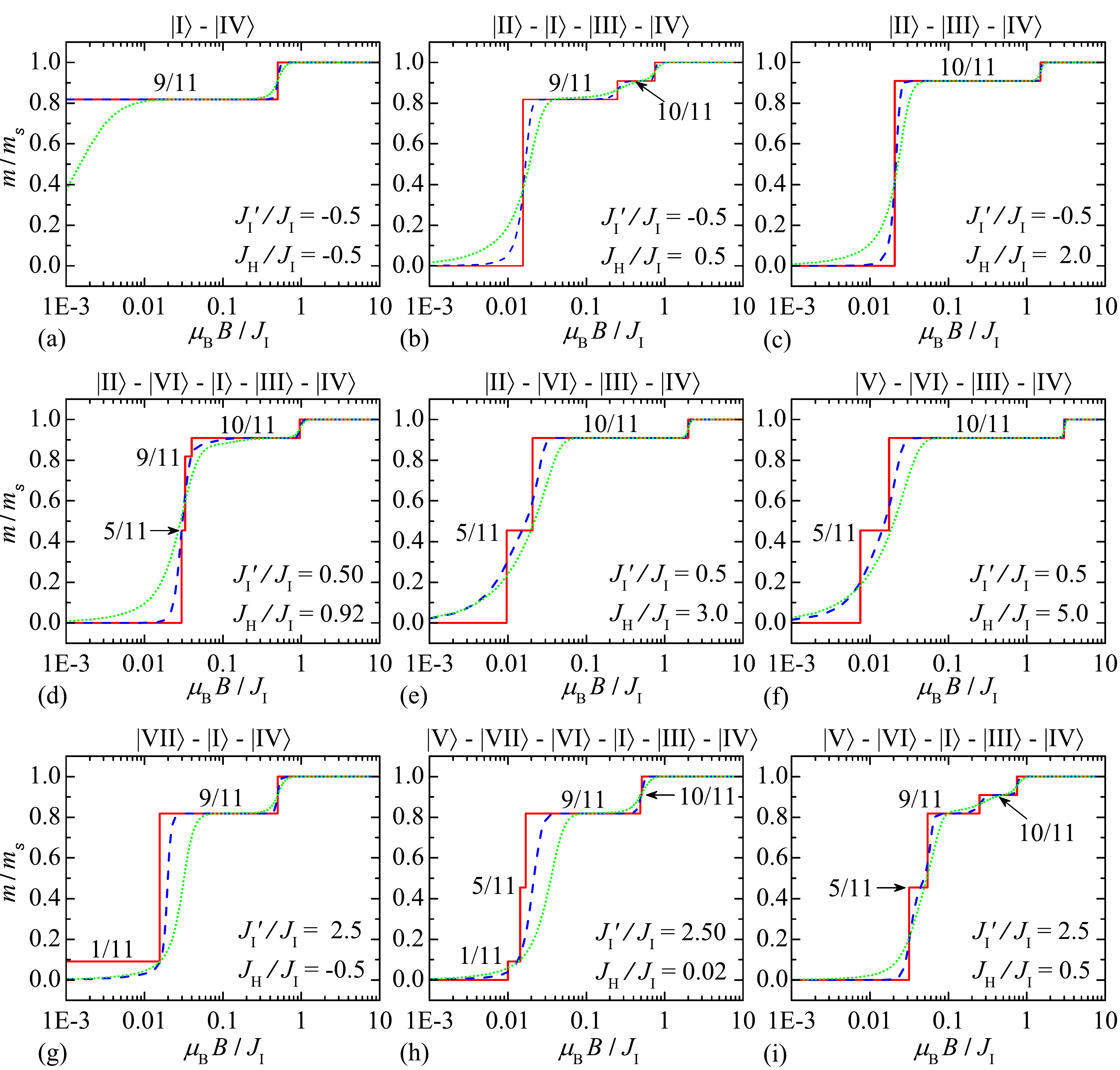}}
\vspace{-0.25cm}
\caption{(Colour online) A semi-logarithmic plot of all representative isothermal magnetization curves of the IH-ODC with the gyromagnetic factors $g_\text{H} = 2$ and $g_\text{I} = 20$ calculated at three different temperatures $k_\text{B}T/J_\text{I} = 0$ (red solid lines), 0.05 (blue dashed lines) and 0.15 (green dotted lines). Different panels demonstrate a diversity of the magnetization process depending basically on a choice of the coupling constants $J_\text{H}/J_\text{I}$ and $J_\text{I}^{\prime}/J_\text{I}$ quoted in the respective panels.} 
\label{fig3}
\end{figure}

The gapped ground states \eqref{eq:I}--\eqref{eq:VII} should be manifested in zero- and low-temperature magnetization curves of the IH-ODC as intermediate plateaus emergent at fractional values of the saturation magnetization. By considering specific values of the gyromagnetic factors $g_\text{H} = 2$ and $g_\text{I} = 20$ one should accordingly detect a zero magnetization plateau in a stability region of the modulated quantum antiferromagnetic phase $|\text{II}\rangle$ and the frustrated quantum antiferromagnetic phase $|\text{V}\rangle$ given by equations~\eqref{eq:II} and \eqref{eq:V}, the intermediate 1/11-plateau may emerge due to the frustrated ferrimagnetic phase $|\text{VII}\rangle$ given by equation \eqref{eq:VII}, the intermediate 5/11-plateau can be ascribed to the modulated quantum ferrimagnetic phase $|\text{VI}\rangle$ given by equation \eqref{eq:VI}, the intermediate 9/11-plateau corresponds to the classical ferrimagnetic phase $|\text{I}\rangle$ given by equation \eqref{eq:I} and finally, the intermediate 10/11-plateau relates to the quantum ferrimagnetic phase $|\text{III}\rangle$ given by equation \eqref{eq:III}. From this perspective, the IH-ODC exhibits a substantial diversity of the magnetization curves with nine possible magnetization scenarios as exemplified in figure~\ref{fig3}.
 
In agreement with the reported ground-state phase diagrams, one finds three different magnetization scenarios with either a single field-driven phase transition $|\text{I}\rangle{-}|\text{IV}\rangle$ [figure~\ref{fig3}~(a)], three field-induced phase transitions $|\text{II}\rangle{-}|\text{I}\rangle{-}|\text{III}\rangle{-}|\text{IV}\rangle$ [figure~\ref{fig3}~(b)] or two field-driven phase transitions $|\text{II}\rangle{-}|\text{III}\rangle{-}|\text{IV}\rangle$ [figure~\ref{fig3}~(c)] on the assumption that the Ising intra-dimer coupling is ferromagnetic $J_\text{I}^{\prime}< 0$. The particular case with a weak antiferromagnetic Ising intra-dimer coupling $J_\text{I}^{\prime}/J_\text{I} \gtrsim 0$ displays  the other three types of magnetization processes due to the presence of the phases $|\text{V}\rangle$ and/or $|\text{VI}\rangle$:  one type involves a sequence of four field-driven phase transitions $|\text{II}\rangle{-}|\text{VI}\rangle{-}|\text{I}\rangle{-}|\text{III}\rangle{-}|\text{IV}\rangle$ [figure~\ref{fig3}~(d)] and the other two types include a sequence of three field-induced phase transitions $|\text{II}\rangle{-}|\text{VI}\rangle{-}|\text{III}\rangle{-}|\text{IV}\rangle$ [figure~\ref{fig3}~(e)] or $|\text{V}\rangle{-}|\text{VI}\rangle{-}|\text{III}\rangle{-}|\text{IV}\rangle$ [figure~\ref{fig3}~(f)], respectively. Finally, the specific case with a sufficiently strong antiferromagnetic Ising intra-dimer coupling $J_\text{I}^{\prime}/J_\text{I} \gg 0$ may exhibit three other  magnetization scenarios, which consecutively include a sequence of two field-driven phase transitions $|\text{VII}\rangle{-}|\text{I}\rangle{-}|\text{IV}\rangle$ [figure~\ref{fig3}~(g)], five field-induced phase transitions $|\text{V}\rangle{-}|\text{VII}\rangle{-}|\text{VI}\rangle{-}|\text{I}\rangle{-}|\text{III}\rangle{-}|\text{IV}\rangle$ [figure~\ref{fig3}~(h)] or four field-induced phase transitions $|\text{V}\rangle{-}|\text{VI}\rangle{-}|\text{I}\rangle{-}|\text{III}\rangle{-}|\text{IV}\rangle$ [figure~\ref{fig3}~(i)], respectively. 

Now, a few comments are in order as far as a thermal stability of the individual magnetization plateaus is concerned. It is quite obvious from figure~\ref{fig3} that some intermediate plateaus are quite robust with respect to thermal fluctuations and they can be clearly discerned in the respective isothermal magnetization curves at low ($k_\text{B}T/J_\text{I} = 0.05$) 
or even at moderate ($k_\text{B}T/J_\text{I} = 0.15$) temperatures. This is typically the case for the zero magnetization plateau and for the intermediate 9/11- and 10/11-plateaus emergent at higher values of  magnetization. On the other hand, the intermediate 1/11- and 5/11-plateaus emergent at smaller values of the magnetization are typically subjected to a considerable smoothing upon increasing the temperature and they cannot be clearly discerned in the respective isothermal magnetization curves even at a relatively low temperature ($k_\text{B}T/J_\text{I} = 0.05$). This striking finding can be related to a substantial difference of the energy gap of the individual ground states ascribed to the relevant magnetization plateaus. As a matter of fact, the gapped ground states, for instance the classical and quantum ferrimagnetic phases $|\text{I}\rangle$ and $|\text{III}\rangle$,  due to their unique nature possess a high energy gap in an excitation spectrum. On the contrary, the highly degenerate ground states, such as the frustrated ferrimagnetic phase $|\text{VII}\rangle$ and the modulated quantum ferrimagnetic phase $|\text{VI}\rangle$, possess only a tiny energy gap and hence, they are subjected to a substantial thermal smoothing owing to their macroscopic degeneracies. 

\section{Magnetization curve of the polymeric compound [Dy$_2$Cu$_2$]$_n$}
\label{sec:4}

\begin{figure}[!t]
\centerline{\includegraphics[width=0.55\textwidth]{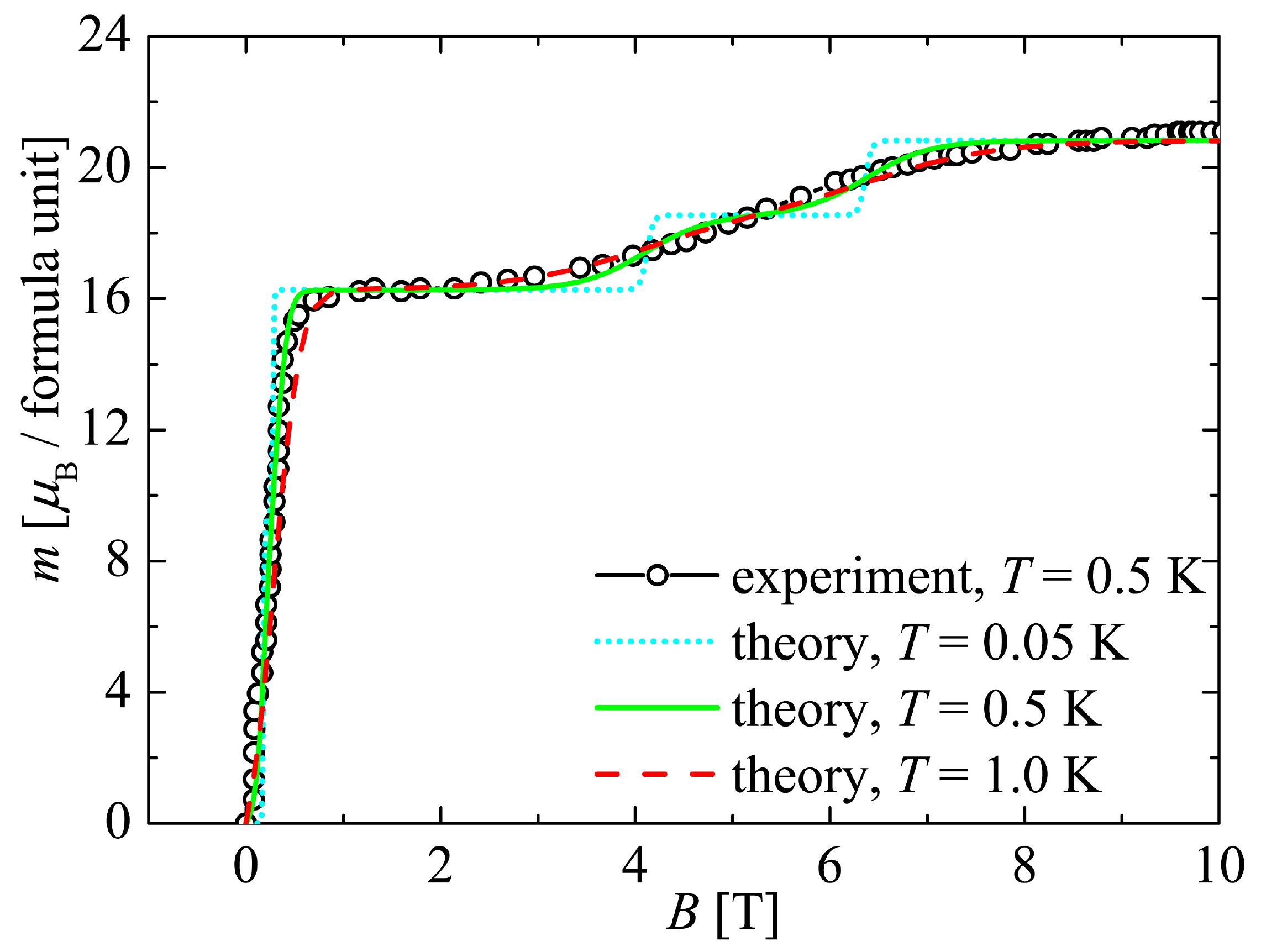}}
\caption{(Colour online) A comparison between the magnetization curve of the polymeric compound [Dy$_2$Cu$_2$]$_n$ measured in pulsed magnetic fields up to 10~T at temperature $T = 0.5$~K (a black solid line with open circles adapted from reference \cite{Oka08}) and the best theoretical fit obtained by using of the IH-ODC with the following fitting set of the parameters: $J_\text{I} / k_\text{B} = 8.02$~K, $J_\text{I}^{\prime} / k_\text{B} = 17.35$~K, $J_\text{H} / k_\text{B} = 1.73$~K, $g_\text{H} = 2.28$ and $g_\text{I} = 18.54$.  The theoretical results for the isothermal magnetization curves are also presented for very low temperature $T = 0.05$~K (light blue dotted line) and slightly higher temperature $T = 1.0$~K (red dashed line) in addition to the temperature $T = 0.5$~K (green solid line) corresponding to the displayed experimental data.} 
\label{fig4}
\end{figure}

In this part we compare available experimental data for the magnetization curve of the 3$d$-4$f$ heterobimetallic coordination polymer [Dy$_2$Cu$_2$]$_n$ measured in pulsed magnetic fields up to 10~T at temperature $T = 0.5$~K \cite{Oka08} with a relevant theoretical prediction based on the IH-ODC. It should be mentioned that the magnetization data recorded in pulsed magnetic fields show a small hysteresis in a low-field region $B \lesssim 0.5$ T, which was ignored for simplicity because it is beyond the scope of the introduced IH-ODC. It actually follows from the inset of figure~4~(a) of reference \cite{Oka08} that the magnetic hysteresis basically depends on a field scan rate and it may be thus attributed to a quantum tunneling of the magnetization typically observed in single-chain magnets due to level crossing among excited states. It is worth noticing that all five model parameters of the IH-ODC defined through the Hamiltonian~\eqref{eq:Htot} were supposed to be free fitting parameters in order to get the best theoretical fit of the experimental magnetization data (see figure~\ref{fig4} for a comparison). The best theoretical fit was accordingly obtained for the following set of the model parameters: $J_\text{I} / k_\text{B} = 8.02$~K, $J_\text{I}^{\prime} / k_\text{B} = 17.35$~K, $J_\text{H} / k_\text{B} = 1.73$~K, $g_\text{H} = 2.28$ and $g_\text{I} = 18.54$. 

The reported value of the gyromagnetic factor $g_\text{H} = 2.28$ is quite typical for Cu$^{2+}$ magnetic ions, while the other reported value $g_\text{I} = 18.54$ is only by a few percent  (cca. 7\%) lower than the value $g_\text{Dy} = 20$ theoretically expected for the effective Ising-spin description of Dy$^{3+}$ magnetic ions with the total angular momentum $J=15/2$ and the respective $g$-factor $g_J = 4/3$ \cite{Jon74,Jen91}. It should be pointed out, moreover, that the values of the Ising inter- and intra-dimer coupling constants $J_\text{I}$ and $J_\text{I}^{\prime}$ should be also rescaled by the factors of 15 and 225 in order to get true values of the coupling constants between Dy$^{3+}$-Cu$^{2+}$ and Dy$^{3+}$-Dy$^{3+}$ magnetic ions when passing away from the effective Ising-spin description of Dy$^{3+}$ magnetic ions. The actual values of the three considered coupling constants consequently read: $J_\text{Dy-Cu} / k_\text{B} = 0.53$~K,  $J_\text{Dy-Dy} / k_\text{B} = 0.08$~K and  $J_\text{Cu-Cu} / k_\text{B} = 1.73$~K, whereas the predicted values of the coupling constants $J_\text{Dy-Cu}$ and $J_\text{Dy-Dy}$ fall into a reasonable range for the coupling constants between 3$d$-4$f$ and 4$f$-4$f$ magnetic ions, respectively \cite{Jon74,Jen91}. Note, furthermore, that the former exchange constant is comparable with the mean value of the exchange coupling $J_\text{Dy-Cu} / k_\text{B} = 0.48$~K, which can be calculated from the coupling constants $J_\text{A} / k_\text{B} = 0.895$~K and $J_\text{B} / k_\text{B} = 0.061$~K assigned previously to two different exchange pathways between Dy$^{3+}$-Cu$^{2+}$ magnetic ions within the simplified tetranuclear model \cite{Oka08}. It is worthwhile to remark that the reported value for the coupling constant $J_\text{Cu-Cu}/ k_\text{B} = 1.73$~K is relatively small with respect to the shortest distance between Cu$^{2+}$-Cu$^{2+}$ magnetic ions, but this small value can be attributed to a rather inefficient exchange pathway involving a double-oxygen bridge between magnetic ($3d_{x^2\text{-}y^2}$) and nonmagnetic ($3d_{z^2}$) orbitals of Cu$^{2+}$ magnetic ions in an elongated square-pyramidal environment [see figure~\ref{fig1}~(a)].     

The predicted values of the coupling constants of the IH-ODC are consistent with the following values of the interaction ratio $J_\text{I}^{\prime}/J_\text{I} \approx 2.2$ and $J_\text{H}/J_\text{I} \approx 0.2$, which should cause a magnetization scenario quite analogous to that shown in figure~\ref{fig3}~(i) with a sequence of four field-driven phase transitions $|\text{V}\rangle{-}|\text{VI}\rangle{-}|\text{I}\rangle{-}|\text{III}\rangle{-}|\text{IV}\rangle$. A low-field part of the magnetization curve should accordingly display two narrow plateaus at zero and approximately 5/11 of the saturation magnetization (note that the gyromagnetic factors slightly deviate from the ideal values $g_\text{H} = 2.0$ and $g_\text{I} = 20.0$), which are, however, completely smeared out by thermal fluctuations at temperatures as low as $T \gtrsim 0.5$~K due to tiny energy gaps of the phases $|\text{V}\rangle$ and $|\text{VI}\rangle$. As a matter of fact, the zero magnetization plateau corresponding to the frustrated quantum antiferromagnetic phase $|\text{V}\rangle$ is restricted to the magnetic fields smaller than $0.17$~T, while the 5/11-plateau related to the modulated quantum ferrimagnetic phase $|\text{VI}\rangle$ is limited to the magnetic-field range $0.17{-}0.28$~T.

On the other hand, the high-field part of the magnetization curve should exhibit two wider intermediate plateaus roughly at 9/11- and 10/11 of the saturation magnetization. The 9/11-plateau pertinent to the classical ferrimagnetic phase $|\text{I}\rangle$ is  stable within the magnetic-field range $0.28{-}4.12$~T, while the 10/11-plateau ascribed to the quantum ferrimagnetic phase $|\text{III}\rangle$ is  stable within the magnetic-field range $4.12{-}6.37$~T. In this regard, it seems quite puzzling that the field-driven transition between the classical and quantum ferrimagnetic phases cannot be evidently seen from the magnetization data recorded for the polymeric compound [Dy$_2$Cu$_2$]$_n$ at a relatively low temperature $T = 0.5$~K as both ground states $|\text{I}\rangle$ and $|\text{III}\rangle$ with a substantial energy gap should be quite resistant with respect to a thermal smoothing. It is plausible to conjecture two possible reasons for the discrepancy between the experimental magnetization data recorded at temperature $T = 0.5$~K and the respective theoretical fit: either the sample of the polymeric compound [Dy$_2$Cu$_2$]$_n$ was not kept during the magnetization process under a perfect isothermal condition since small heating (e.g. due to the magnetocaloric effect) could resolve this discrepancy as evidenced by a theoretical curve calculated for a slightly higher temperature $T = 1.0$~K, or the discrepancy is of intrinsic origin and it comes from two different exchange pathways between Dy$^{3+}$-Cu$^{2+}$ magnetic ions neglected within the investigated quantum spin chain.     

\section{Conclusion}
\label{sec:5}

In the present article we have scrupulously investigated diverse ground-state phase diagrams and magnetization curves of the IH-ODC by assuming two different gyromagnetic factors of the Ising and Heisenberg spins. The investigated quantum spin chain including two different Land\'e $g$-factors was inspired by the polymeric coordination compound [Dy$_2$Cu$_2$]$_n$, whose magnetic structure shows a peculiar one-dimensional architecture with regularly alternating dimeric units of Dy$^{3+}$-Dy$^{3+}$ and  Cu$^{2+}$-Cu$^{2+}$ magnetic ions stacked in an orthogonal fashion \cite{Oka08}. The vertical dimer of Dy$^{3+}$-Dy$^{3+}$ magnetic ions was approximated by a couple of the Ising spins, while the horizontal dimer of Cu$^{2+}$-Cu$^{2+}$ magnetic ions was approximated by a couple of the Heisenberg spins.  

It has been found that the IH-ODC exhibits a rich variety of the classical and quantum ground states, which, besides the fully saturated paramagnetic phase emergent at sufficiently high magnetic fields, involve  six more ground states: the frustrated and modulated quantum antiferromagnetic phase, the frustrated and modulated quantum ferrimagnetic phase, as well as, the quantum and classical ferrimagnetic phase. These remarkable ground states are responsible for the presence of magnetization plateaus in zero- and low-temperature magnetization curves, which are manifested at 0, 1/11, 5/11, 9/11 and/or 10/11 of the saturation magnetization. A stability of the intermediate magnetization plateaus with respect to thermal fluctuations was investigated in detail. It has been verified that the rather narrow 1/11- and 5/11-plateaus ascribed to the frustrated and modulated quantum ferrimagnetic phases with a high macroscopic degeneracy and a tiny energy gap are easily destroyed upon a small increase of temperature, while the relatively wide 9/11- and 10/11-plateaus ascribed to the nondegenerate classical and quantum ferrimagnetic phases with a robust energy gap are quite resistant with respect to thermal fluctuations and may be also discernible  at relatively low (up to moderate) temperatures.

Consequently, we have successfully applied the IH-ODC with two different gyromagnetic factors of the Ising and Heisenberg spins  for a theoretical modelling of the high-field magnetization data recorded previously for the polymeric coordination compound [Dy$_2$Cu$_2$]$_n$ at a sufficiently low temperature $T = 0.5$~K~\cite{Oka08}. The respective low-temperature magnetization curve evidently displays intermediate magnetization plateau(s), which starts at a relatively high value of the magnetization being approximately 78\% of the saturated value. The best theoretical fit of the available experimental data based on the IH-ODC suggests that the magnetization plateau(s) observed experimentally could be ascribed to the classical and quantum ferrimagnetic phases given by equations \eqref{eq:I} and \eqref{eq:III} inherent to the intermediate 9/11- and 10/11-plateau, respectively. It should be mentioned, however, that the magnetization data measured for the polymeric complex [Dy$_2$Cu$_2$]$_n$ at low enough temperature $T = 0.5$~K do not provide any experimental evidence for the relevant field-induced phase transition between the classical and quantum ferrimagnetic phase \cite{Oka08}. This discrepancy could be resolved either by a small deviation from a perfect isothermal condition during the experimental measurement caused by heating of the sample (e.g., due to the magnetocaloric effect) or it may relate to the oversimplified nature of the introduced IH-ODC neglecting two structurally inequivalent exchange pathways between Dy$^{3+}$-Cu$^{2+}$ magnetic ions. In our future work we plan to examine the magnetization process of the IH-ODC taking into consideration two different exchange constants between Dy$^{3+}$-Cu$^{2+}$ magnetic ions in order to clarify this issue.  

\section*{Acknowledgements}
The authors are deeply indebted to Oleg Derzhko for a lot of insightful and inspiring scientific discussions, from which they have substantially benefited over their whole careers.  
This work was financially supported by Slovak Research and Development Agency provided under the contract No.~APVV-16-0186 and by The Ministry of Education, Science, Research and Sport
of the Slovak Republic provided under the grant No. VEGA 1/0105/20.

\newpage
\ukrainianpart

\title{Пояснення природи плато намагнічення 3d-4f координаційного полімера [Dy$_2$Cu$_2$]$_n$ на основі спін-1/2 ортогонально-димерного ланцюжка Ізінґа-Гайзенберґа}
\author{Й. Стречка \refaddr{upjs}, Л. Ґалісова \refaddr{tuke}, Т. Верхоляк \refaddr{icmp}}
\addresses{
\addr{upjs} Інститут фізики, Факультет природничих наук, Університет імені П. Й. Шафарика, парк Ангелінум 9, Кошиці 04001, Словаччина
\addr{tuke} Інститут управління виробництвом, Факультет виробничих технологій у Прешові, \\ 
            Технічний університет в Кошицях, вул. Баєрова 1, Прешов 08001, Словаччина
\addr{icmp} Інститут фізики конденсованих систем НАН України, вул. Свєнціцького, 1,
	79011 Львів, Україна
}

\makeukrtitle

\begin{abstract}
Детально досліджено основний стан і процес намагнічення точно-розв'язного спін-$1/2$ ортогонально-димерного ланцюжка Ізінґа-Гайзенберґа з двома різними гіромагнітними факторами Ізінґових та Гайзенберґових спінів. Показано, що досліджений квантовий спіновий ланцюжок виявляє до семи можливих основних станів в залежності від взаємної дії магнітного поля, констант внутрішньо- і між-димерної взаємодії. А саме, фрустрована та модульована антиферомагнітні фази відповідальні за нульове плато намагнічення при нульовій температурі, проміжні 1/11- та 5/11-плато виникають у фрустрованій та модульованій квантових ферімагнітних фазах, в той час як проміжні 9/11- та 10/11-плато можна віднести до квантової і класичної ферімагнітних фаз. Запропоновано, що плато намагнічення експериментально спостережені при високих полях у 3d-4f гетеробіметалічному координаційному полімері [\{Dy(hfac)$_2$(CH$_3$OH)\}$_2$\{Cu(dmg)(Hdmg)\}$_2$]$_n$ (H$_2$dmg = dimethylglyoxime; Hhfac = 1,1,1,5,5,5-hexafluoropentane-2,4-dione) можна віднести до класичної та квантової феромагнітних фаз.

\keywords ортогонально-димерний ланцюжок Ізінґа-Гайзенберґа, плато намагнічення, 3d-4f координаційний полімер
\end{abstract}

\end{document}